\documentclass[prd,eqsecnum,twocolumn,amsfonts,showpacs]{revtex4}
\usepackage{bm}
\def\fsl#1{\setbox0=\hbox{$#1$}           
   \dimen0=\wd0                                 
   \setbox1=\hbox{/} \dimen1=\wd1               
   \ifdim\dimen0>\dimen1                        
      \rlap{\hbox to \dimen0{\hfil/\hfil}}      
      #1                                        
   \else                                        
      \rlap{\hbox to \dimen1{\hfil$#1$\hfil}}   
      /                                         
   \fi}                                         %
\newcommand{\be}{\begin{equation}}
\newcommand{\ee}{\end{equation}}
\newcommand{\bea}{\begin{eqnarray}}
\newcommand{\eea}{\end{eqnarray}}
\newcommand{\beq}{\begin{equation}}
\newcommand{\eeq}{\end{equation}}
\newcommand{\beqs}{\begin{eqnarray}}
\newcommand{\eeqs}{\end{eqnarray}}

\begin{document}

\title{Gauge-Invariant Quantities Characterizing Gauge Fields in 
Chromodynamics} 

\author{Gouranga C. Nayak$^{(a,b)}$ and Robert Shrock$^a$}

\affiliation{
(a) \ C.N. Yang Institute for Theoretical Physics \\
State University of New York \\
Stony Brook, NY 11794}

\affiliation{
(b) \ Department of Physics \\
University of Illinois at Chicago \\
Chicago, IL 60607}

\begin{abstract}

We calculate Lorentz-invariant and gauge-invariant quantities characterizing
the product $\sum_a D_R(T^a) F^a_{\mu\nu}$, where $D_R(T^a)$ denotes the matrix
for the generator $T^a$ in the representation $R=$ fundamental and adjoint, for
color SU(3).  We also present analogous results for an SU(2) gauge theory.

\end{abstract}


\maketitle

\newpage

\pagestyle{plain}
\pagenumbering{arabic}

\section{Introduction}
\label{intro}

Although the properties and interactions of quarks and gluons require for their
description a quantum field theory, quantum chromodynamics (QCD), it has proved
useful to consider the semiclassical limit of this theory in certain cases.
For example, successful models of high-energy particle production and
hadronization have made use of a non-abelian Yang-Mills generalization of the
Schwinger mechanism \cite{schwinger,he} in which the chromoelectric field
inside a flux tube between an initial quark-antiquark pair is responsible for
subsequent nonperturbative production of $q \bar q$ pairs and hadronization
\cite{cnn}-\cite{pythia}. The Schwinger calculation itself described the
nonperturbative production of a charged fermion-antifermion pair by a constant
classical electric field, the result of which can also be obtained from the
imaginary part of the Euler-Heisenberg effective action \cite{he}.  The
semiclassical limit of chromoelectric fields has also been used in certain
models of relativistic heavy ion collisions \cite{qgp1}-\cite{ilm}. For the
case of a classical SU(3) gauge field that is constant in space and time and is
such that the chromomagnetic field vanishes and all group components of the
chromoelectric field point in the same direction (e.g., ${\bf E}^a = E^a \hat z
\ \forall \ a$), general formulas for the nonperturbative production of gluon
pairs $gg$ and $q \bar q$ pairs have recently been given \cite{gg,qqbar}.

Proceeding from the special case of static, spatially constant classical fields
to the general case of spacetime-dependent classical fields, one recalls that
Euclidean solutions of classical non-abelian gauge theories with with
nontrivial topological index, i.e., instantons, have played an important role
in understanding the properties of these theories
\cite{bpst,hooft,cdg,shuryak}. In particular, analyses of semiclassical
effects due to instantons have shown that, in the case of weak SU(2)$_L$, these
lead to nonperturbative violation of $B$ and $L$ (conserving $B-L$)
\cite{hooft} and may be significant for baryogenesis at the electroweak phase
transition \cite{eb}, while in the color SU(3) case, these analyses of
instanton effects have explained, among other things, the breaking of the
global axial vector isoscalar U(1)$_A$ symmetry and hence the fact that the
$\eta'$ meson is not an almost Nambu-Goldstone boson \cite{u1a}.  Classical
solutions have also been relevant for classification of Yang-Mills theories
(mainly in the SU(2) case) \cite{wuyang}-\cite{early}.

Given this importance of semiclassical color fields, it seems useful to have a
set of gauge-invariant quantities that characterize these fields.  Accordingly,
in this paper, we present such a set.  We consider an SU($N$) gauge theory,
concentrating on the case of color, $N=N_c=3$, but also giving some results for
the simpler case $N=2$.  We calculate certain gauge-invariant and
Lorentz-invariant quantities that characterize the product
\beq
({\cal F}_R)_{\mu\nu} \equiv \sum_a D_R(T^a) F^a_{\mu\nu} \ , 
\label{fr}
\eeq
where $D_R(T^a)$ denotes the matrix for the generator $T^a \equiv T_a$ of
SU($N$) in the representation $R$, a sum over the group index $a$ from 1 to
$N^2-1$ is understood, and we consider the case of $R$ being the fundamental
and adjoint representation.  The dimension of the representation $R$ is denoted
$d_R$.  We recall that for the fundamental representation,
$[D_{fund}(T^a)]_{ij}=(T^a)_{ij}$, $1 \le i,j \le N$ and for the adjoint,
$[D_{adj}(T^a)]_{bc} = -ic_{abc}$, $1 \le a,b,c \le N^2-1$, where the structure
constants $c_{abc}$ of the SU($N$) Lie algebra are defined via
$[T^a,T^b]=ic_{abc}T^c$ with normalization determined by the standard condition
${\rm Tr}(T^aT^b)=(1/2)\delta^{ab}$.  We also recall the relation, for SU($N$),
$\{T_a,T_b\}=(1/N)\delta_{ab} \cdot 1_{N \times N} + d_{abc}T_c$.  The field
strength tensor is $F^a_{\mu\nu}=\partial_\mu A^a_\nu - \partial_\nu A^a_\mu +
gc_{abc}A^b_\mu A^c_\nu$, where $g$ is the gauge coupling (taken positive
without loss of generality).

One may contrast the way in which the results of calculations are expressed in
terms of gauge-invariant quantities in classical and quantum field theory.  In
perturbative quantum field theory calculations involving internal gauge boson
lines, this entails the cancellation of the associated gauge parameter between
different Feynman diagrams contributing to the amplitude for a given
process. In a nonperturbative quantum field theory calculation of some
gauge-invariant operator ${\cal O}$, one actually performs the average over the
gauge fields in the path integral, e.g., in the widely used lattice gauge
theory fomulation,
\beq
\langle {\cal O} \rangle = \frac{\int [\prod_{n,\mu} dU_{n,\mu} d\psi_n 
d\bar\psi_n]   \, {\cal O} \, e^{-S}}
{\int [\prod_{n,\mu} dU_{n,\mu} d\psi_n d\bar\psi_n] \, e^{-S}}
\label{oave}
\eeq
where $S$ denotes the (Euclidean) action and both the measure and action are
gauge invariant.  For example, in pure gluodynamics with a Euclidean action $S
= - \beta \sum_{plaq.} (1/N){\rm Re}[{\rm Tr}_f(U_{plaq.})]$ where Tr$_f$ is
the trace in the fundamental representation, 
$U_{plaq.}$ denotes the product of $U's$ around a
plaquette, and $\beta=2N/g_0^2$, a strong-coupling expansion of a glueball mass
would be conveniently expressed in a series in $\beta$ or, equivalently, as a
character expansion.  The situation in a (semi)classical gauge theory
calculation is different from either of these types of calculations in quantum
field theory, since it depends directly on the field strengths. This was
already evident from the Schwinger calculation of the production of a
fermion-antifermion pair by an electric field ${\bf E}$ that is constant in
space and time, namely,
\beq
\frac{dW}{d^4x}= \frac{(q e E)^2}{4\pi^2}\sum_{n=1}^\infty 
n^{-2} e^{-n \pi m^2/(|q|eE)}
\label{schw}
\eeq
where $q$ denotes the charge of the fermion.  We recall how this is expressed
in terms of Lorentz-invariant and gauge-invariant quantities.  In this abelian
case the field strength tensor $F_{\mu\nu}=\partial_\mu A_\nu - \partial_\mu
A_\nu$ itself is gauge-invariant, in contrast to the non-abelian case.
Particle production occurs only if $|{\bf E}| > |{\bf B}|$, and in this case
one can transform to an inertial frame in which the magnetic field is zero,
whence the result in eq. (\ref{schw}).

This calculation was generalized recently to the non-abelian color group
SU(3)$_c$ in the special case in which (i) there is only a chromoelectric
field, ${\bf E}^a$, i.e., the chromomagnetic field ${\bf B}^a=0$, (ii) ${\bf
E}^a$ is a constant in space and time, and (iii) all of the group components of
${\bf E}^a$ point along the same spatial direction.  In this case the
production rates for gluon pairs $gg$ and quark-antiquark pairs $q \bar q$
\cite{gg,qqbar} were calculated.  For example, for $q \bar q$ it was found that
\beq
\frac{dW_{q \bar q}}{d^4x~d^2p_T} = -\frac{1}{4\pi^3} 
\sum_{j=1}^3 |g\lambda_{q,j}| \ln 
[1-e^{-\pi(p_T^2+m^2)/|g\lambda_{q,j}|}]
\label{qqbarpt}
\eeq
where ${\bf p}_T$ denotes the momentum of the quark transverse to the direction
of the chromoelectric field ${\bf E}^a = E^a \hat z$ and 
where the $\lambda_{q,j}$ depends on two gauge-invariant, Lorentz-invariant
quantities 
\beq
C_1 = \sum_{a} (E^a)^2
\label{c1}
\eeq
and
\beq
C_2 = \Big (\sum_{a,b,c} d_{abc}E^aE^bE^c \Big )^2 \ , 
\label{c2}
\eeq
where the sums of SU(3)$_c$ group indices $a,b,c$ are from 1 to 8. 
Integration over ${\bf p}_T$ yields 
\beq
\frac{dW_{q \bar q}}{d^4x} = \frac{1}{4\pi^2} 
\sum_{j=1}^3 (g\lambda_{q,j})^2 
\sum_{n=1}^\infty n^{-2} e^{-n \pi m_q^2/|g\lambda_{q,j}|}
\label{qqbar}
\eeq
As was noted by Schwinger \cite{schwinger}, it is necessary to take account of
the renormalization of the gauge coupling in the presence of a constant
electric field, and the same is true for the non-abelian case. Thus, strictly
speaking, where we write $e$ or $g$, these refer to running couplings, which
run as a function of (invariants of) the respective gauge fields.

\section{Generalities on Quantities Characterizing $({\cal F}_R)_{\mu\nu}$} 

We now proceed to analyze the general case where both a non-abelian electric
and magnetic field are present and where neither is a constant in space or
time.  Under a (local) SU($N$) gauge transformation generated by the unitary
matrix $U$,
\beq
({\cal F}_R)_{\mu\nu} \to D_R(U) ({\cal F}_R)_{\mu\nu} D_R(U)^{-1}
\label{ftran}
\eeq
where $D_R (U^{-1})=[D_R(U)]^{-1}$. $({\cal F}_R)_{\mu\nu}$ is a matrix (of
dimension $d_R \times d_R$) in group space. Given that $({\cal F}_R)_{\mu\nu}$
transforms as in eq. (\ref{ftran}), it follows that the characteristic
polynomial equation for ${\cal F}_R)_{\mu\nu}$ is invariant under a gauge
transformation, and hence so are its roots, the eigenvalues.

Since we will carry out various matrix manipulations with the field strength
tensor, it will be convenient to use the pseudo-Euclidean metric, in which
there is no distinction between covariant and contravariant indices.
In this case, with the ordering of the indices given by $x^\mu = ({\bf x},it)$,
the field strength tensor takes the form 
\beq
F^a_{\mu\nu} = \left [
\begin{array}{cccc}
0 & B^a_3 & -B^a_2 & -iE^a_1  \\
-B^a_3 & 0 & B^a_1 & -iE^a_2  \\
B^a_2 & -B^a_1 & 0 & -iE^a_3  \\
iE^a_1 & iE^a_2 & iE^a_3 & 0 \end{array} \right ] 
\label{famunu}
\eeq
for $a=1,...N^2-1$.  The dual field strength tensor is then $\tilde
F^a_{\mu\nu} = (i/2)\epsilon_{\mu\nu\rho\sigma}F_{\rho\sigma}$, where
$\epsilon_{\mu\nu\rho\sigma}$ is totally antisymmetric, with
$\epsilon_{1234}=1$.  For each group index $a$, $F^a_{\mu\nu}$ changes via a
similarity transformation under a (homogeneous) Lorentz transformation ${\cal
U}$, viz.,
\beq
F^a \to {\cal U} F^a {\cal U}^{-1} 
\label{flor}
\eeq
in a notation suppressing explicit Lorentz indices.  Therefore, the
characteristic polynomial equation for this matrix, and its roots, are
Lorentz-invariant.  For an individual $a$, these eigenvalues are not
gauge-invariant, but they will be useful at intermediate steps in our
calculation of the gauge-invariant quantities characterizing $({\cal
F}_R)_{\mu\nu}$.  The eigenvalues are determined from the characteristic
polynomial equation
\beq
{\rm det}(F^a - \lambda \cdot 1)=0 \ . 
\label{fcharpoly}
\eeq
where here $1$ is the $4 \times 4$ identity matrix.  We use the following
relations, which hold individually for each group index $a$:
\beq
{\rm Tr}_{Lor.}[(F^a)^2] = F^a_{\mu \nu} F^a_{\nu\mu} = 
2(|{\bf E}^a|^2-|{\bf B}^a|^2)
\label{trF2}
\eeq
and
\beqs
{\rm Tr_{Lor.}} [(F^a)^4] & \equiv & 
F^a_{\mu\nu}F^a_{\nu\rho}F^a_{\rho\sigma}F^a_{\sigma\mu} \cr\cr
& = &  2(|{\bf E}^a|^2-|{\bf B}^a|^2)^2 + 4({\bf E}^a \cdot {\bf B}^a)^2 \ . 
\cr\cr
& & 
\label{trF4}
\eeqs
(Note also that ${\rm Tr}_{Lor.}(F^a\tilde F^a) = F_{\mu\nu}\tilde F_{\nu\mu}
=-4{\bf E}^a \cdot {\bf B}^a$.)  With these inputs, the characteristic
polynomial equation takes the form, for each $a$,
\beq
(\lambda^a)^4 - (|{\bf E}^a|^2-|{\bf B}^a|^2)(\lambda^a)^2 - 
({\bf E}^a \cdot {\bf B}^a)^2=0 \ . 
\label{fcharpoly2}
\eeq
The solutions are
\beq
\lambda^a_1 = -\lambda^a_3 = \sqrt{x^a_1}
\label{lam13}
\eeq
\beq
\lambda^a_2 = -\lambda^a_4 = \sqrt{x^a_2}
\label{lam24}
\eeq
where
\beqs
x^a_{1,2} & = & \frac{1}{2}\bigg [ |{\bf E}^a|^2-|{\bf B}^a|^2 \cr\cr
& & \pm \Big [ (|{\bf E}^a|^2-|{\bf B}^a|^2)^2 + 
4({\bf E}^a \cdot {\bf B}^a)^2 \Big ]^{1/2} \ \bigg ] \cr\cr
& = & \frac{1}{4} \bigg [ F^a_{\mu\nu} F^a_{\nu\mu} \pm 
\Big [(F^a_{\mu\nu} F^a_{\nu\mu})^2 + (F^a_{\mu\nu}\tilde F^a_{\nu\mu})^2
 \Big ]^{1/2} \ \bigg ] \cr\cr
& & 
\label{x12}
\eeqs
Although a parity or time reversal transformation flips the sign of
$F^a_{\mu\nu}{\tilde F}^a_{\nu\mu}$, it leaves the $x^a_{1,2}$ invariant
since they depend on $F^a_{\mu\nu}{\tilde F}^a_{\nu\mu}$ only via its 
square. 

For the general SU($N$) case we define 
\beq
C_{k1} = \sum_a (\lambda_k^a)^2
\label{ck1}
\eeq
and
\beq
C_{k2}=\Big [\sum_{a,b,c} d_{abc}\lambda_k^a \lambda_k^b \lambda_k^c 
\Big ]^2 \ .
\label{ck2}
\eeq
where the sums over the SU($N$) group indices run over
$a,b,c=1,...,N^2-1$, and
\beq
r_k \equiv \frac{3C_{k2}}{{(C_{k1}})^3} \ . 
\label{rk}
\eeq
These quantities will be used below.

\section{Invariants for $({\cal F}_R)_{\mu\nu}$: General Method for SU($N$)} 

We next use these Lorentz-invariant eigenvalues $\lambda_k^a$ to calculate the
gauge-invariant quantities characterizing $({\cal F}_R)_{\mu\nu}$.  For a $d_R
\times d_R$ dimensional matrix $A$ in group space we denote ${\rm Tr}_R(A)
\equiv \sum_{i=1}^{d_R} A_{ii}$.  Taking the trace over group indices and
Lorentz indices, we have 
\beqs
{\rm Tr_R}[{\rm Tr_{Lor.}} h({\cal F}_R)] & = & 
\sum_{k=1}^4 \ {\rm Tr_R}\{h(T^a \lambda_k^a)\} \cr\cr
& = & \sum_{k=1}^4 {\rm Tr_R} ( h(V_k)) \cr\cr
& = & \sum_{k=1}^4 \sum_{\ell=1}^{d_R} h(\Lambda_{k\ell})
\label{eigen}
\eeqs
where $\Lambda_{k\ell}$ for $\ell=1,...,d_R$ are the eigenvalues of
\beq
(V_k)_{ij} \equiv \sum_{a} [D_R(T^a)]_{ij} \lambda_k^a \ , \quad 
1 \le i,j \le d_R
\label{vk}
\eeq
for each $k$.  To find the gauge-invariant and Lorentz-invariant quantities
characterizing $({\cal F}_R)_{\mu\nu}$, one can evaluate the group traces of
the the set of matrices $V_k$, $V_k^2$, ..., $V_k^{d_R}$.

\section{Invariants for $({\cal F}_I)_{\mu\nu}$ for SU(2)}

We label the representations of SU(2) by an isospin, $T$, taking on integral or
half-integral values. The single diagonal generator has the form $T^3={\rm
diag}(-I,-I+1,..., I-1,I)$, so that the components of a representation are 
$|I,I_3\rangle$ satisfying $T^2|I,I_3\rangle = I(I+1) \, |I,I_3\rangle$ and 
$T_3|I,I_3\rangle = I_3 \, $, with $T_3 |I,I_3\rangle$. 
Applying our procedure, we find that for this theory,
$\sum_a D_I(T^a)F^a_{\mu\nu}$ is chacterized by the invariants
\beq
\Lambda_{k\ell} = I_\ell C_{k1}
\label{su2inv}
\eeq
where $I_\ell = I_3$, with $C_{k1}$ evaluated for $N=2$ in
eq. (\ref{ck1}). $\Lambda_{k\ell}$ does not depend on $C_{k2}$ since 
$d_{abc}=0$ for SU(2).

\section{Invariants for $({\cal F}_f)_{\mu\nu}$ in SU(3)$_c$} 

We first consider a function $h$ that has a Taylor series expansion in powers
of $(({\cal F})_R)_{\mu\nu}$; we then apply this for the case where $R$ is the
fundamental ($f$) representation of SU(3)$_c$, for which we need the traces
over group indices of $V_k$, $V^2_k$, and $V^3_k$.  We find, for each $k$,
\beq
\sum_{\ell=1}^3 \Lambda_{k\ell} = 0 
\label{lambdaeq1}
\eeq
\beq
\sum_{\ell=1}^3 (\Lambda_{k\ell})^2 = \frac{1}{2}\sum_{a} (\lambda_k^a)^2
\label{lambdaeq2}
\eeq
\beq
\sum_{\ell=1}^3 (\Lambda_{k\ell})^3 =
\frac{1}{4} \, \sum_{a,b,c} d_{abc}\lambda_k^a \lambda_k^b 
\lambda_k^c \ . 
\label{cas}
\eeq
For a particular $k$, the solution of the above equations is 
follows:
\beq
\Lambda_{k\ell} = \sqrt{\frac{C_{k1}}{3}} \, 
\cos \Big (\theta_k + \frac{2(\ell-1)\pi}{3} \Big ) \ , \quad \ell=1,2,3 
\label{eigenf}
\eeq
where $\theta_k$ is given by 
\beq
\cos^2(3\theta_k) = r_k 
\label{cssq}
\eeq
and here $C_{k1}$ and $C_{k2}$ are given by eqs. (\ref{ck1}) and (\ref{ck2})
evaluated for $N=3$, and $r_k$ by eq. (\ref{rk}).  (Note that $0 \le
3C_{k2}/C_{k1}^3 \le 1$.)  Because of the symmetries $\lambda^a_1=-\lambda^a_3$
and $\lambda^a_2=-\lambda^a_4$, there are thus 4 independent invariants here,
which can be taken to be $C_{k1}$ and $C_{k2}$ for $k=1,2$.  For the case of
identically zero chromomagnetic field, ${\bf B}^a=0 \ \forall \ a$,
$C_{21}=C_{22}=0$, while $C_{11}$ and $C_{12}$ reduce to the quantities denoted
$C_1$ and $C_2$ in eqs. (\ref{c1}) and (\ref{c2}).

\section{Invariants for $({\cal F}_{adj})_{\mu\nu}$} 

We next calculate the invariants for $({\cal F}_{adj})_{\mu\nu}$ using the
relation $[D_{adj}(T^a)]_{bc}=-ic_{abc}$.  Again, we focus on the case of
color, $N=3$, setting $c_{abc}=f_{abc}$, and first evaluate the determinant
\beqs
& & Det[f^{abc}\lambda_k^a -\Lambda\delta_{bc}] =  
\Lambda^2[ \Lambda^6+{\cal A}_k\Lambda^4+{\cal B}_k\Lambda^2+{\cal C}_k ] 
\cr\cr
& = & 
\Lambda^2\Pi_{\ell=1}^3(\Lambda-i\Lambda_{k\ell})(\Lambda+i\Lambda_{k\ell}) \ .
\label{eg1}
\eeqs
Since $f^{abc}\lambda_k^c\lambda_k^a=0$, it follows that $\lambda_k^a$ is an
eigenvector of the matrix $(V_k)^{ab}=f^{abc}\lambda^c_k$ with zero eigenvalue.
Since for $N=3$, $V^{ab}_k=f^{abc}\lambda_k^c$ is an even-dimensional real 
antisymmetric matrix, its eigenvalues (i) are comprised of opposite-sign pairs,
and (ii) are pure imaginary (so the eigenvalues of $-if^{abc}\lambda_k^c$,
which are the $\Lambda_{k\ell}$, are real), whence 
\beqs
& & [D_{adj}(T^a)_{bc}\lambda_k^c]_{eigenvalues} = \cr\cr
& &(\Lambda_{k1}, \Lambda_{k2}, \Lambda_{k3},0,-\Lambda_{k1}, -\Lambda_{k2}, 
-\Lambda_{k3},0).
\label{ffa}
\eeqs
The coefficients ${\cal A}_{k,n}$ of $\Lambda^n$ in eq. (\ref{eg1}) are 
${\cal A}_{k,8}=1$ and 
\beq
{\cal A}_k=\frac{3}{2}C_{1k} 
\label{ak}
\eeq
\beq
{\cal B}_k=\frac{9}{16}C_{1k}^2 
\label{bk}
\eeq
and
\beq
{\cal C}_k = \frac{C_{1k}^3}{16}(1-r_k) \ . 
\label{ck}
\eeq
From eq. (\ref{eg1}) we find
\beq
\sum_{\ell=1}^3 \Lambda_{k\ell}^2={\cal A}_k
\label{akleq1}
\eeq
\beq
\Lambda_{k1}^2\Lambda_{k2}^2 +\Lambda_{k2}^2\Lambda_{k3}^2+
\Lambda_{k3}^2\Lambda_{k1}^2={\cal B}_k
\label{akleq2}
\eeq
and
\beq
\Lambda_{k1}^2\Lambda_{k2}^2\Lambda_{k3}^2={\cal C}_k \ . 
\label{akleq3}
\eeq
We define
\beq
\cos(3\phi_k)=2r_k-1 \ . 
\label{thetaf}
\eeq
The solution of these three equations is 
\beq
\Lambda_{k\ell} = \bigg [ \frac{C_{k1}}{2}\Big \{ \cos \Big ( \phi_k +
  \frac{2(\ell-1)\pi}{3} \Big ) \Big \} \bigg ]^{1/2}
\label{eigena}
\eeq
for $\ell=1,2,3$. Again, owing to the symmetries (\ref{lam13}) and
(\ref{lam24}), these eigenvalues depend on four functionally independent
invariants, which may be taken to be $C_{k1}$ and $C_{k2}$ for $k=1,2$. 
We note that an equivalent set of solutions for the $\Lambda_{k\ell}$ is 
\beqs
& & \bigg [ \frac{C_{k1}}{2}(1-\cos\theta'_k) \bigg ]^{1/2}; 
 \cr\cr
& & \bigg [ \frac{C_{k1}}{2}
\Big [ 1+ \cos \Big (\frac{\pi}{3} \pm \theta'_k \Big ) \Big ] \bigg ]^{1/2}
\label{eigenaold}
\eeqs
where \cite{mp}
\beq
\cos(3\theta'_k) = -1+2r_k \ . 
\label{thetaprime}
\eeq

\section{Some Further Remarks}

We add here some further remarks.  First, since the invariants for $({\cal
F}_f)_{\mu \nu}$ and $({\cal F}_{adj})_{\mu \nu}$ depend on the same set of
four independent invariants, they clearly can be related to each other.  We
note parenthetically that the Schwinger mechanism for pair production in an
electric field was generalized to an expression for pair production in an
oscillatory electric field in Ref. \cite{itzyk,fried} (see also
\cite{dunne}). Expressions for $dW/d^4x$ were given for the case of a general
time-dependent electric field and also for the SU(3)$_c$ case with general
time-dependent ${\bf E}^a=E^a(t) \hat z$ in Ref. \cite{nayak1}.  Our set of
invariants may be used for the still more general problem of nonperturbative
particle production for spacetime-dependent classical gauge fields.

\section{Discussion and Conclusion}

In this paper we have given a general method for calculating the
gauge-invariant and Lorentz-invariant quantities characterizing the products
$\sum_a D_R(T^a) F^a_{\mu\nu}$ for an SU($N$) gauge group.  We have applied our
method to compute these quantities for all representations of SU(2) and for the
fundamental and adjoint representations of SU(3).  Our results apply for the
most general case of spacetime-dependent gauge fields and can provide a
convenient set of quantities in terms of which to express calculations for
classical chromodynamics.

\acknowledgments

This work was supported in part by the YITP National Science Foundation grants
NSF-PHY-03-45822 and NSF-PHY-06-53342.  The Univ. of Illinois at Chicago
address for G. N. applies after Nov. 16, 2007.

\end{document}